\documentclass[preprint2]{aastex}
 \usepackage{epsfig}
\usepackage{natbib}
\newcommand{\lb}{$\cal L_{B}$ }

\shorttitle{A model for the flares of Sagittarius A*}
\shortauthors{Tagger and Melia}

\begin{document}
\title{A Possible Rossby Wave Instability Origin for the Flares in Sagittarius A*}

\author{Michel Tagger }
\affil{Service d'Astrophysique, (UMR AstroParticules et Cosmologie), CEA Saclay \\ 91191 Gif-sur-Yvette, France}
\email{tagger@cea.fr}
\author{Fulvio Melia$^1$}
\affil{Departments of Physics and Astronomy, The University of Arizona,\\
Tucson AZ 85721, USA}
\email{fmelia@as.arizona.edu}
\altaffiltext{1}{Sir Thomas Lyle Fellow and Miegunyah Fellow.}

\begin{abstract}
In recent years, near-IR and X-ray flares have been detected from the Galaxy's central 
radio point source, Sagittarius A* (Sgr A*), believed to be a $\sim 3\times 10^6\;M_\odot$ supermassive
black hole. In some cases, the transient emission appears to be modulated with a (quasi-)periodic
oscillation (QPO) of $\sim 17-20$ minutes. The implied $\sim 3\,r_S$ size of the emitter (where $r_S\equiv
2GM/c^2$ is the Schwarzschild radius) points to an instability---possibly induced by accretion---near 
the Marginally Stable Orbit (MSO) of a slowly spinning object. But Sgr A* is not accreting 
via a large, `standard' disk; instead, the low density environment surrounding it apparently 
feeds the black hole with low angular momentum clumps of plasma that circularize within 
$\sim 10-300\,r_S$ and merge onto a compact, hot disk. In this {\it Letter}, we follow the 
evolution of the disk following such an event, and show that
a Rossby wave instability, particularly in its magnetohydrodynamic (MHD) form, grows rapidly 
and produces a period of enhanced accretion lasting several hours. Both the amplitude of
this response, and its duration, match the observed flare characteristics rather well. 
\end{abstract}
\section{Introduction}
The prospect of identifying phenomena within several Schwarzschild radii ($r_S\equiv 2GM/c^2$)
of Sgr A* has improved considerably in recent years, with the detection of near-IR \citep{GSO03}
and X-ray \citep{BGM05} flares modulated with an average period of $\approx 17-20$ minutes
\citep{BGM06}.  This (quasi-)period corresponds to a modulation near the MSO of the black hole 
\citep{MBLW01,LM02}, possibly suggesting a transient event associated with a 
wave pattern co-rotating with the gas, or perhaps a ``hot'' spot where matter has fallen 
in from larger radii, impacting the small disk \citep{FM97}.  In this {\it Letter}, we 
examine numerically the Rossby wave instability (RWI) induced 
by a clump of magnetized plasma merging onto the disk and focus on its implications 
for the subsequent flow of matter through the MSO toward the event horizon. 

For a black hole mass $M\sim 3\times 10^6\; M_\odot$, the radius $r\approx 3\,r_S$ inferred 
from the average period is interesting for several reasons.  Aside from  
constraining the flaring event to a region very near the MSO, it supports 
the view that there is indeed a hot, magnetized disk confined to the inner $\sim 10r_S$ that 
can simultaneously produce a $\approx 10\%$ polarized component and a position angle that 
rotates by $90^\circ$ across the mm/sub-mm bump \citep{BML01}. In addition, strong variability 
is seen near the middle of these events, in which the X-ray flux drops by a factor of 40--50\ \% in 
10--15 minutes. Simple light travel time arguments constrain the emitting region to be no bigger 
than $\sim 17-34\,r_S$ \citep{M01}. This tight orbit 
suggests that the X-ray and mm/sub-mm photons are produced by the same medium, probably via 
synchrotron-self-Compton processes.
Finally, 
this is in line with the measured intrinsic size of Sgr A* at $\lambda$7 mm \citep{B04}.  

Extensive hydrodynamic \citep{MC99,CN05} and MHD \citep{IN02} simulations
show that for the stellar-wind fed conditions at the Galactic center, the average specific 
angular momentum of gas captured gravitationally by Sgr A* is too small to sustain a 
`conventional' (i.e., typically large $\sim 10^5\,r_S$) disk.  Instead, only clumps of plasma 
with relatively small angular momentum venture inwards and merge with---essentially, `rain' 
onto---the compact disk at the circularization radius, which for Sgr A* is $\la
10-10^3\,r_S$ (see also Melia \& Falcke 2001, Melia 2006). In this paper, we model the
transient disruption to the disk induced by blobs of plasma circularizing within the inner
$\sim 10-20\,r_S$, and show that the ensuing dynamical evolution has characteristics in common 
with those inferred from the flares. 

\section{Magnetic Rossby Wave Instability}\label{sec:RWI}
The instability we are here concerned with has a long history, dating back to \cite{LH78}, who 
showed that a disk presenting an extremum of a quantity $\cal L$ (later dubbed vortensity) was 
subject to a local instability of Rossby vortices. The requirement of an extremum is similar to 
that giving rise to the Kelvin-Helmholtz instability of sheared flows.
More recently, \cite{LLC99}Ê (see also Li et al. 2000, Li et al. 2001) renamed it the Rossby 
Wave Instability (RWI) and developed the theory and numerical simulation.  \cite{VT05} used it to address the 
presence of a {\lq Dead Zone\rq} in protostellar disks. 

In isothermal, unmagnetized disks, $\cal L$ is the specific vorticity averaged across 
the disk thickness, 
\begin{equation}
{\cal L}\ =\ \frac{\vec\nabla \times \vec V}{\Sigma}\ =\ \frac{\kappa^2}{2\Omega\Sigma}\;,
\end{equation}
where $\Sigma$ is the disk's surface density, $\Omega$ its rotation frequency, and 
\(
\kappa^2\ =\ 4\Omega^2+2\Omega\Omega'r
\)
is the epicyclic frequency squared.
The extremum of $\cal L$ appears to be due to an extremum in the radial density 
profile. However, starting from a very different point of view---the seismology of relativistic 
accretion disks---\cite{NW91} described g-modes that intoduce an extremum in $\kappa$, and thus 
also in $\cal L$ (though this time due to the relativistic rotation curve), near the MSO (where 
$\kappa$  vanishes).
%
The g-mode of \citeauthor{NW91} is thus similar to the RWI, but they seem to have missed the 
mechanism that makes it unstable. Its tentative association with a 
QPO in X-ray binaries does not seem to be supported by observations. 
We will see, however, that it plays an important role in what follows.

Many of the details regarding how the instability is driven have already appeared in the work 
cited above. The main idea is that Rossby waves in disks form normal modes 
trapped near the extremum of $\cal L$. The waves propagate either outside or inside their 
corotation radius, i.e., the radius where the wave rotates at the same frequency as the gas, 
depending on the sign of the gradient in $\cal L$  \citep{T01}. They have a positive energy in 
the first case, and a negative energy in the second. 
In a forthcoming paper, we shall discuss the MHD form of the RWI, for a disk threaded by a 
vertical (poloidal) magnetic field $B_{0}(r)$. Its properties are essentially the same as 
those discussed above, except that here the critical quantity is 
\(
{\cal L}_{B}={\kappa^2\Sigma}/({2\Omega}{B_{0}^2}),
\)
and the growth rate can be higher because of the long-range 
action of the Lorentz force coupling the Rossby vortices \citep{TP99}. Another difference is that 
the energy and angular momentum of the Rossby waves can escape the disk vertically as Alfv\'en waves 
propagating along the magnetic field lines \citep{VT02}. 

Let us finally mention that even without an extremum in $\cal L$ or $\cal L_{B}$, normal modes may 
still occur if the disk has an inner edge that reflects spiral density waves. In magnetized disks this gives the Accretion-Ejection Instability \citep{TP99}. Recently, \cite{LGN03} discussed the non-axisymmetric 
g-modes and concluded that those without nodes in their vertical structure 
(that we analyze here) are not unstable. But this conclusion, based on the absence of a reflecting 
inner edge in a relativistic disk, neglected the possibility of trapped modes based on Rossby, rather 
than density, waves.

\section{Numerical simulations}
\begin{figure}[t] 
\plotone{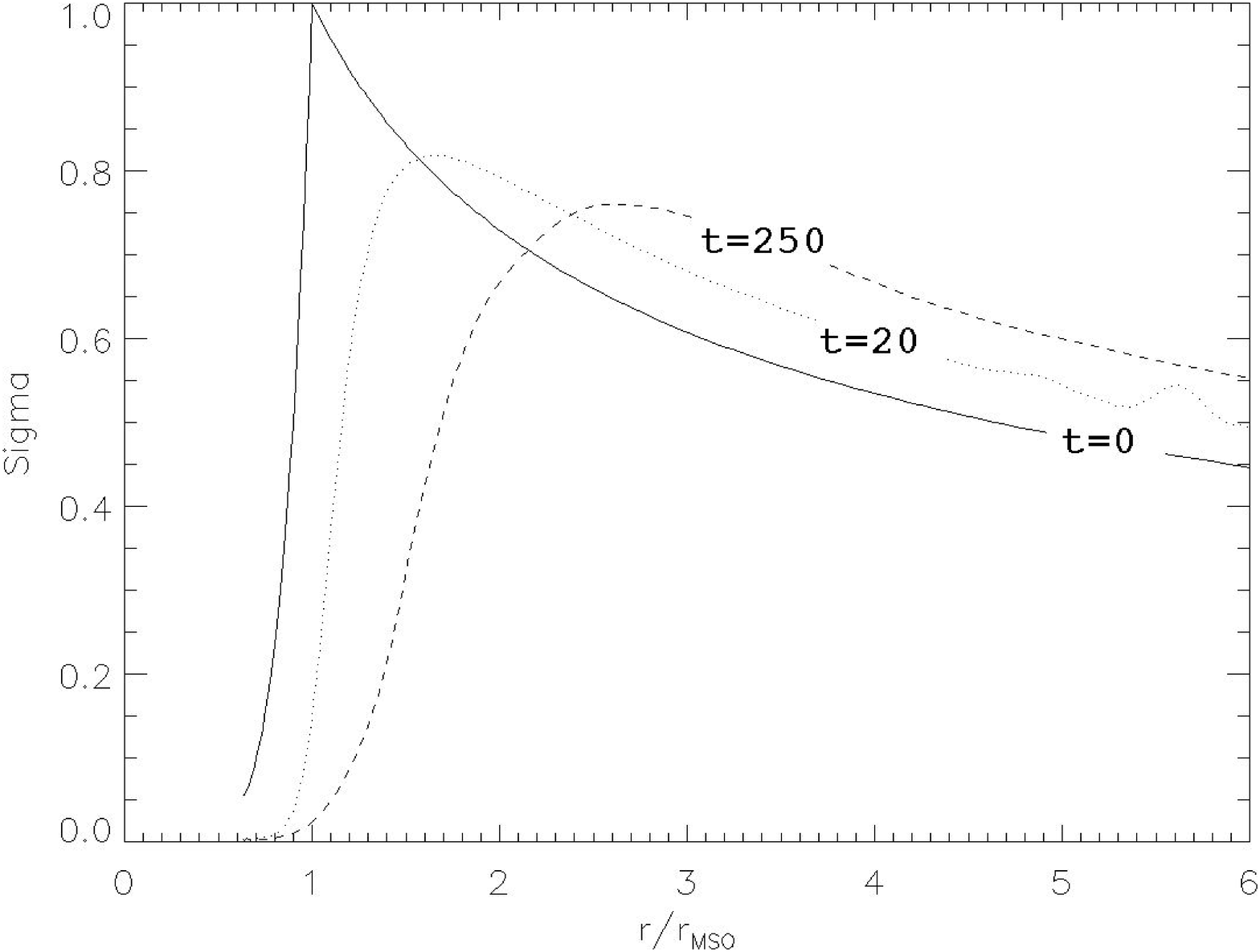}
\plotone{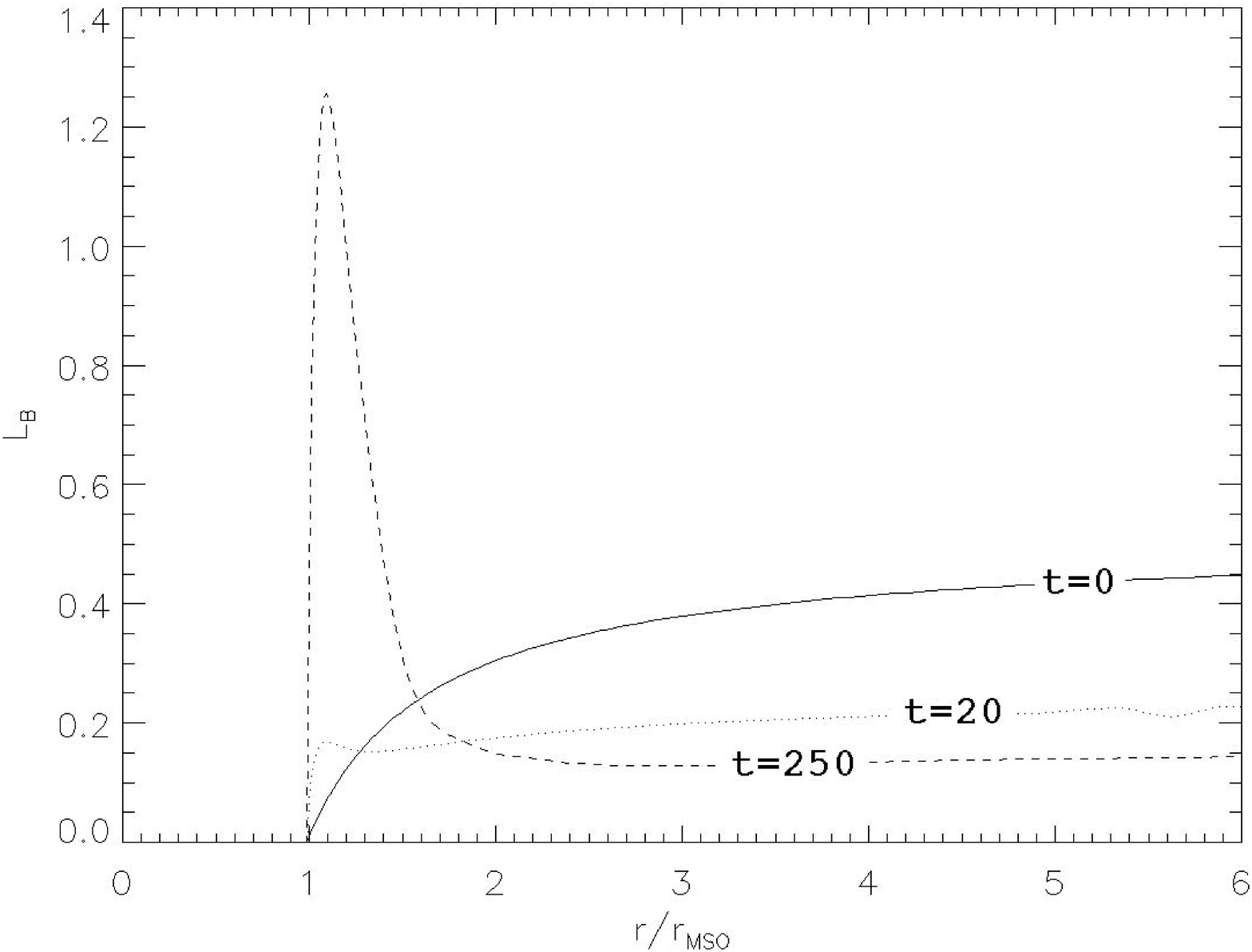}
\caption{Surface density (top) and \lb (bottom) profiles, at startup (full) and after 20 
(dashed) and 250 (dot-dashed) MSO orbital times  in our first simulation. Radii are scaled to 
$r_{MSO}$.}
   \label{fig:nobumpSigma}
\end{figure}
The simulations presented here are relevant to the conditions in Sgr A*. We use the 2-D 
MHD code introduced in \cite{CT01}, that considers an infinitely thin disk in vacuum. This is well adapted 
to the physics we consider, since the RWI and spiral modes have a structure that is essentially constant 
across the disk thickness, allowing us to integrate the dynamical equations over $z$. (On the other hand, 
this simplification forbids us to study the Magneto-Rotational Instability (MRI), whose structure does vary 
with $z$.)  

The radial grid points are distributed logarithmically, allowing simulations that extend far in radius 
(typically with $r_{max}/r_{min}\approx 50$), so as to minimize unwanted boundary effects.  We have 
modified the code by including the pseudo-Newtonian potential of Paczinsky-Witta, 
\(
\Phi(r)={GM}/({r-r_{S}}).
\)
This allows for the existence of an MSO, $r_{MSO}=3r_{S}$, where $\kappa$ vanishes. Furthermore, 
at the inner grid point the gas is allowed to cross freely during its final collapse toward the central object. 
The simulations are carried out on a polar grid with $n_{r}=256$ and $n_{\theta}=128$ points, and extend 
from $\sim 0.6$ to $30\ r_{MSO}$. The MSO is at the 30th radial grid point, allowing us to resolve 
properly the gas flow (including the sonic point) in the plunging region; the infall velocity at the 
inner boundary is typically 3 to 5 times the sound velocity. Since pressure does not play a major role 
in the physics of Rossby waves, we fix the sound speed as a function of radius, which for simplicity
is kept constant throughout the simulations.  We first present the simulation of a disk with initially 
a power-law surface density profile, $\Sigma\sim r^{-1/2}$, and a sound speed $c_{S}=.05\,r \Omega(r)$, 
(giving a constant disk aspect ratio $h/r$). The  initial magnetic 
field is such that $\beta\equiv8\pi p/B^2=2$ throughout the disk. After a few 
orbits, the density profile near the MSO reaches a quasi-steady state, as shown in 
Figure~\ref{fig:nobumpSigma} (here and below we scale all radii to $r_{MSO}$ and times to the orbital 
period there). This is very similar to what is obtained with 3-D codes \citep{HK01, MM03}. An m=1 perturbation, 
identified as a g-mode instability by the RWI mechanism, sets in, gradually eroding the inner edge of the 
disk (since there is no turbulent accretion from farther out). This builds up the extremum of \lb just beyond 
the MSO. The instability has its corotation radius near the extremum, as expected. The accretion rate $\dot M$
through the disk peaks when the instability gets established, and then decreases gradually for the rest of the simulation, 
as the density maximum slowly moves outward.

\begin{figure}[t] 
\plotone{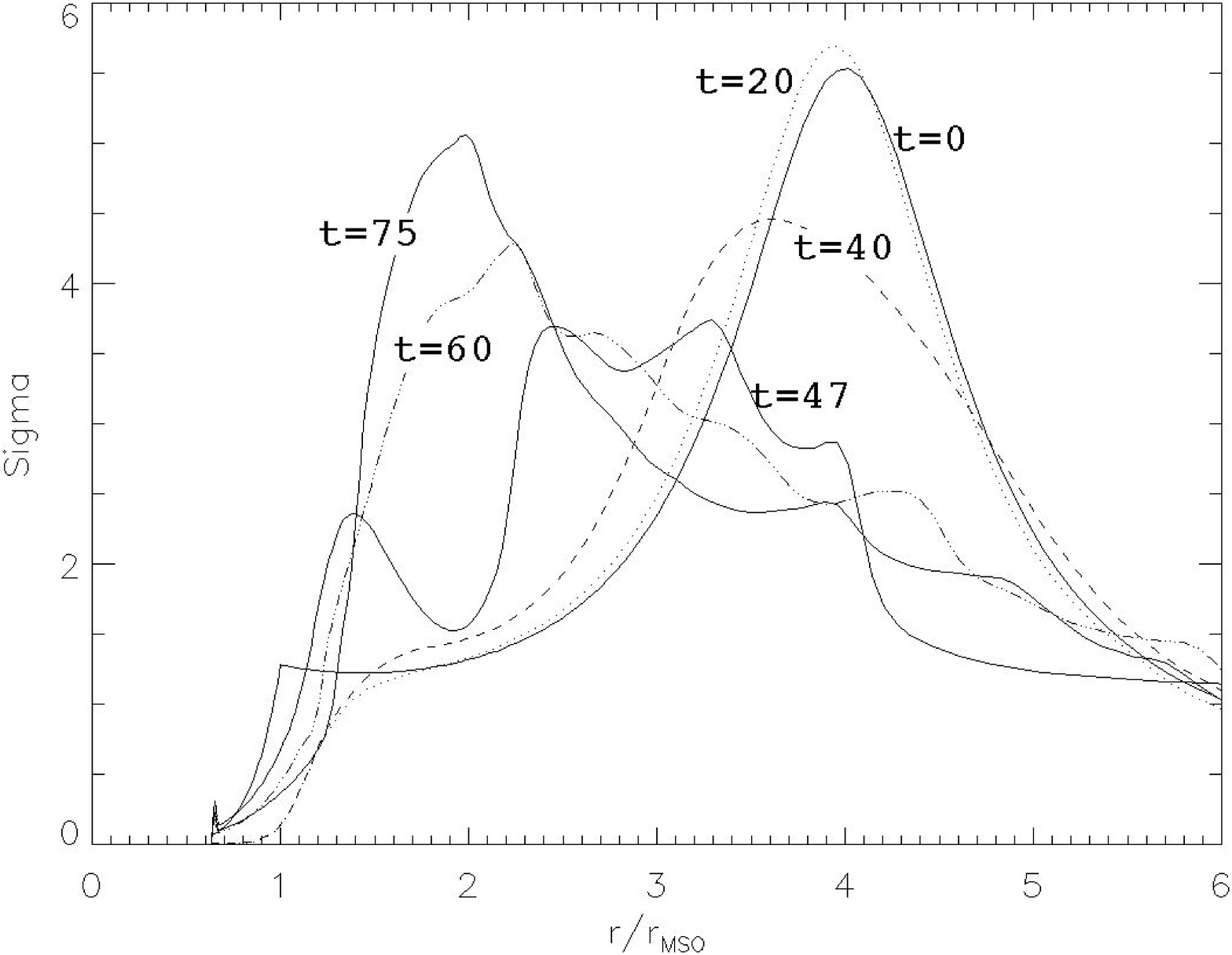} 
\plotone{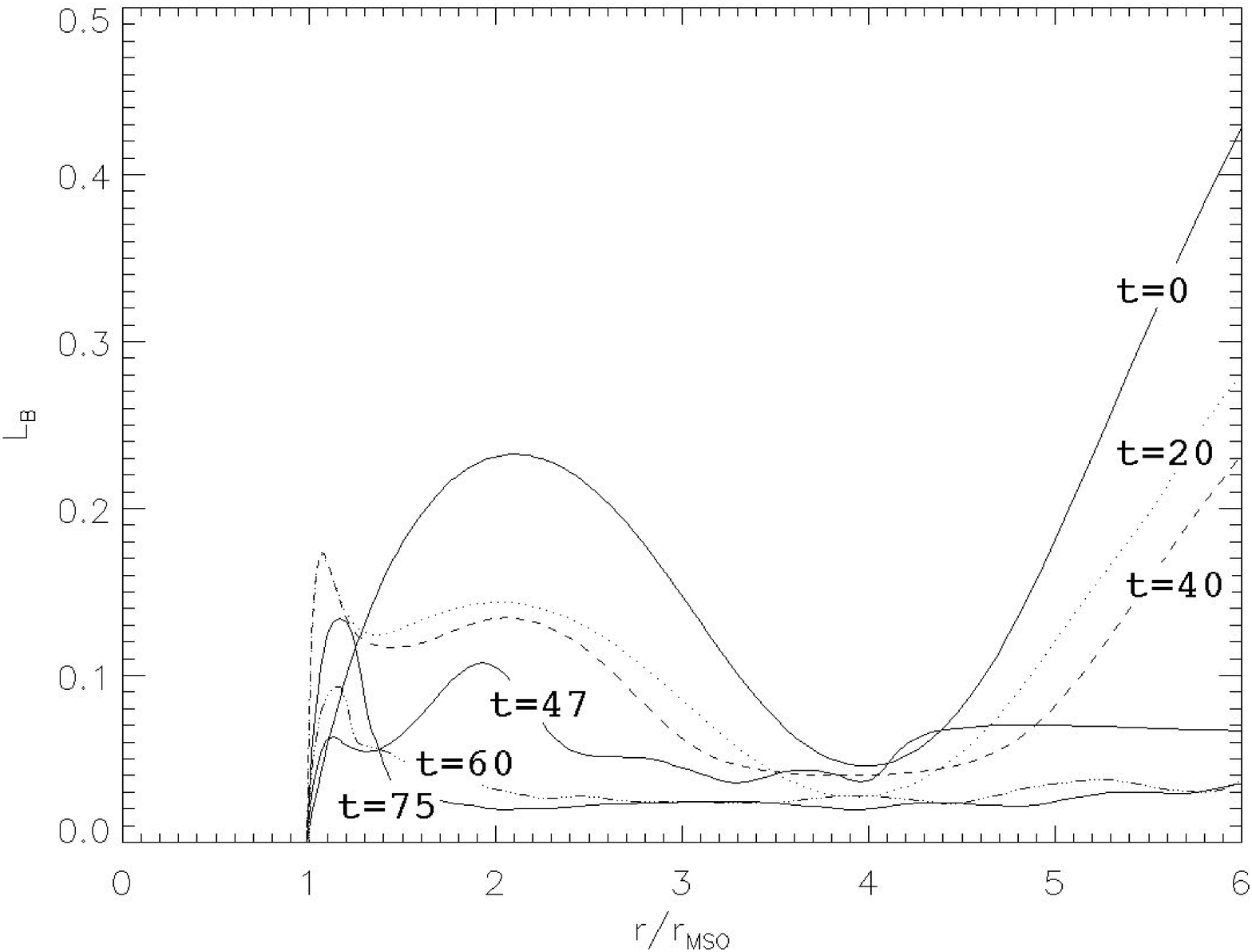} 
\caption{Surface density (top) and \lb (bottom) profiles, at startup (full) and after 20 , 40, 47, 60, 75  
MSO orbital times during our second simulation.}
   \label{fig:bump4Sigma}
\end{figure}

In our second calculation, we simulate a flare event by adding to the `standard' disk a strong bump 
in density and magnetic field centered at $4\ r_{MSO}$, where it generates a minimum in $\cal L_{B}$. 
This bump represents 56\% of the total mass in the central $6 \ r_{MSO}$.  Pursuant to our discussion 
in the introduction, we assume that a blob of magnetized plasma with very little angular momentum has 
merged onto the disk, rapidly circularizing to form a ring at this radius.  As shown in Figure~\ref{fig:bump4Sigma}, 
the evolution is much more violent. In a nearly sudden episode between 40 and 60 MSO orbital times (hardly 
2 periods at $4r_{MSO}$, or in real time about $5$ hours), most of the added gas has been transported to the 
inner region of the disk by violent instabilities developping over the whole region. 

\begin{figure}[t] 
\centering
 \plotone{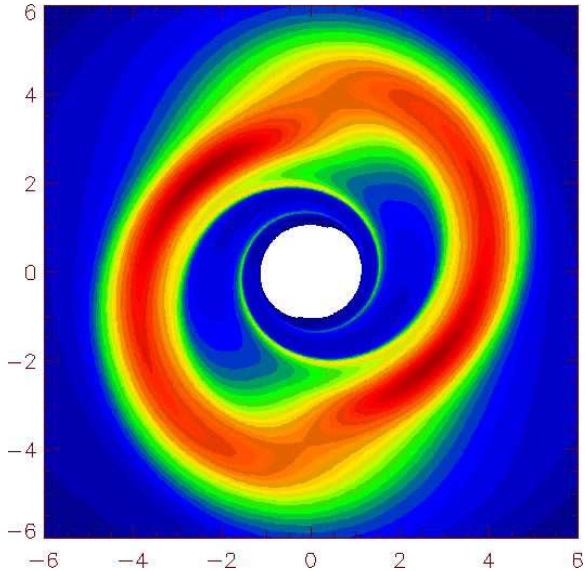}
   \caption{Surface density in our second simulation at $t=40\ t_{MSO}$, when the bump at 
$4\ r_{MSO}$ (red, dark) starts being disrupted by the instability.}
   \label{fig:rho40}
\end{figure}

We see $m=1$ modes developing (as previously) near the inner edge, and $m=2$ in the bump region, with a  
corotation near the extremum of $\cal L_{B}$. These modes appear to be non-linearly coupled; similar coupling 
has been observed in simulations and observations of spiral galaxies \citep{TSA87, MT97}. This coupling can 
be explosive (i.e., result in faster than exponential growth) when it involves modes of positive and negative 
energy \citep{DRR69}.  Figure~\ref{fig:rho40} shows the surface density profile at $t=40t_{MSO}$, 
when the density bump starts being disrupted by the instability. Figure~\ref{fig:bump4Sigma} shows 
that after this rapid phase, the radial profile of \lb becomes perfectly flat beyond the MSO region,  
confirming qualitatively that it is that profile to which the instability reacts.
\begin{figure}[t] 
 \plotone{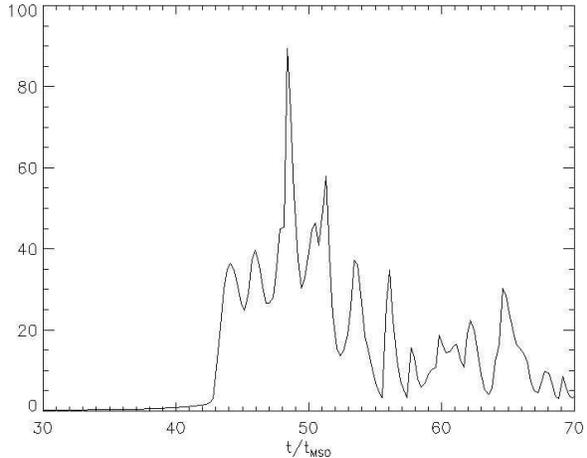}
   \caption{Accretion rate through the inner edge of the simulation box in the second simulation.}
   \label{fig:lightcurve}
\end{figure}

Producing realistic light curves involves much more physics 
than we include here, and is beyond the scope of the present simulations. However, 
figure~\ref{fig:lightcurve} shows $\dot M$ through the inner edge of the simulation box. 
Notice the brief episode of strong, enhanced accretion between $t=40t_{MSO}$ and $t=60t_{MSO}$, with
the largest amplitude occurring over a time interval of roughly $7t_{MSO}$ (i.e., $\sim 140$ minutes).
The similarity between this time and the duration of a typical flare ($\sim 2-3$ hours) is very 
encouraging. Also, there is clear evidence of 4 or 5 quasi-periodic pulsations during this
phase of enhanced accretion, mimicking the behavior witnessed during several IR flares and the
XMM event of 2004. This appears to be due to an axisymmetric g-mode, trapped near the extremum 
of \lb and non-linearly excited by the instabilities. The details may vary with the numerical setup, 
e.g., with the amount of artificial viscosity; this is reasonable since this is the only dissipation 
limiting the growth of the instability in the simulations (besides the destruction of the bump itself), 
but the main features (duration of the flare and pulsations) remain the same.

Simulations of less magnetized or unmagnetized disks, or of a weaker bump, give a very similar, though 
somewhat slower, evolution. Placing the bump at 6 or 8 $r_{MSO}$ delays the development of the instability, 
but does not change the duration or the amplitude of the \lq flare\rq, which occurs only when the mass of 
the accreted gas approaches the MSO. However this might result in changes of the mm and cm emissivity
relative to the IR/X-ray emission, so a comparative study of the broadband spectrum of many flares
may provide indirect evidence of the variation in clumping radius.  
\section{Conclusion}

The RWI, particularly in its MHD form, can produce the behavior required to account for the near-IR 
and X-ray flares in Sgr A*. Unstable both near the MSO and in the annulus where a clump of interstellar 
plasma has merged onto the disk, it develops as strong Rossby vortices whose non-linear evolution can 
be remarkably fast. The instability grows in a few rotation times, and causes the clump to accrete to 
the inner edge of the disk. The resulting period of enhanced $\dot M$ at the inner edge corresponds to 
the final stage of this collapse, so that its duration (comparable with the observed duration of the flares) 
does not depend on the initial location of the clump. Furthermore, the enhanced $\dot M$ is accompanied 
by QPO very similar to those observed both in IR and in X-rays. The quasi-period reflects the underlying 
Keplerian frequency near the MSO, though the exact quasi-period may be dependent on our use of a 
pseudo-Newtonian potential, rather than a fully relativistic model. Natural extensions will thus be to 
include the metric of a spinning Black Hole, and to consider the relevance of the RWI to explain the 
high-frequency QPO in microquasars.

While current wisdom seems to be settling on an accretion instability for the origin of
Sgr A*'s flares, the process by which the actual emission occurs is uncertain. 
%
%
The favored scenario right now connecting the mm/sub-mm to NIR and X-rays portions of the spectrum 
\citep{E05, LM01} is that the former is due to synchrotron, whereas the X-rays are produced by 
synchrotron-self-Compton (SSC). It turns out that producing the right blend of physical conditions to 
fit both the NIR and X-ray flare emission (under the assumption that the two occur more or less 
simultaneously) is not trivial. 

In a related paper \citep{LMP05}, we considered a scenario in which the radiative manifestation 
of the flare occurs when the energy released by the accretion instability is transferred to a 
population of nonthermal particles via stochastic acceleration by the turbulent component of 
the magnetic field.  Indeed, as mentioned in \S \ref{sec:RWI}, Rossby waves in the disk extend 
into the corona as Alfv\'en waves, carrying much of the accretion energy. Clearly, much work 
remains to be done in coupling these complementary approaches.  One of the main results reported 
in this paper is that the enhanced $\dot M$ during a flare event is consistent---both in amplitude 
and temporal evolution---with the characteristics observed in Sgr A*'s flares. We are in the process 
of self-consistently calculating the particle acceleration following an RWI-induced event, and will 
report the results of this work in a future publication.

This research was partially supported by NSF grant AST-0402502 (at Arizona). FM is grateful to the
College de France, where this work was carried out. MT wishes to thank P. Varni\`ere for help with 
the numerical simulations. FM and MT thank G. B\'elanger for stimulating discussions.

\bibliographystyle{apj}
\bibliography{GCBHbib}

\end{document}